# Magnetocaloric effect of $Fe_{47.5}Ni_{37.5}Mn_{15}$ bulk and nanoparticles: A cost-efficient alloy for room temperature magnetic refrigeration

Chang-Gi Lee[a], Varatharaja Nallathambi[b,c], TaeHyeok Kang[d], Leonardo Shoji Aota[c], Sven Reichenberger[b], Ayman El-Zoka[e], Pyuck-Pa Choi[d], Baptiste Gault[c,e], Se-Ho Kim[a,c,*]

[a] Department of Materials Science & Engineering, Korea University, Seoul, 02841, Republic of Korea

[b] Technical Chemistry I and Center for Nanointegration Duisburg-Essen (CENIDE), University of Duisburg-Essen, 45141 Essen, Germany

[c] Max-Planck-Institut für Eisenforschung GmbH, Max-Planck-Straße 1, 40237 Düsseldorf, Germany

[d] Department of Materials Science & Engineering, Korea Advanced Institute of Science and Technology, Daejeon, 34141, Republic of Korea

[e] Faculty of Engineering, Department of Materials, Imperial College London, London, SW7 2AZ, United Kingdom

*corresponding author: sehonetkr@korea.ac.kr

## Table of contents (TOC)

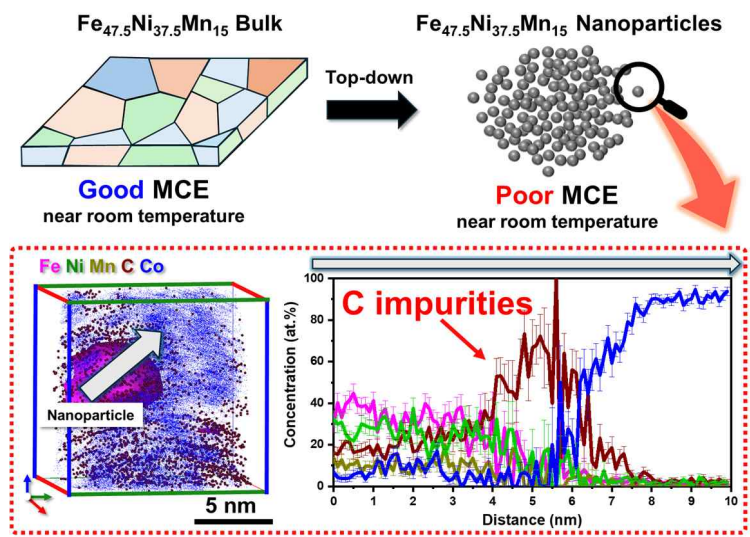

**Abstract**

The development of magnetic refrigerators that operate at room temperature without the use of environmentally harmful substances represents a significant advancement in eco-friendly technology. These refrigerators employ the magnetocaloric effect (MCE), which has traditionally been achieved using expensive rare-earth elements such as gadolinium. To facilitate cost-effective commercialization, it is essential to investigate alternative materials, such as transition metal alloys. In this study, an $Fe_{47.5}Ni_{37.5}Mn_{15}$ alloy, which has a Curie temperature that is close to room temperature, is cast, and the alloy exhibits a noteworthy cooling power of 297.68 J/kg, which makes it good for cost-effective applications. To further enhance MCE performance through super-paramagnetism (e.g. size reduction), nanoparticles of the same composition are synthesized using a top-down approach via pulsed-laser ablation in ethanol. However, these nanoparticles do not exhibit a Curie temperature near room temperature, likely due to significant carbon incorporation during synthesis, which adversely affected their magnetocaloric properties. This study underscores the potential of transition metal alloys for magnetic refrigeration and highlights the need for optimized synthesis methods to achieve desired thermal properties in nanoparticulate form.

**Introduction**

The magnetocaloric effect (MCE) is essential for the development of next-generation, energy-efficient cooling systems, such as magnetic refrigeration operating at room temperature. This phenomenon is intrinsically linked to the concept of entropy within material systems. When a magnetic field is applied to a magnetic material in adiabatic condition, the alignment of magnetic moments leads to a decrease in magnetic entropy, while the total entropy of the system should remain constant, provoking an increase in temperature. The interplay between magnetic field and materials can offer a path to modulate temperature, which can be indirectly estimated based on the thermodynamic equations (see supplementary information).[1–4] This sensitivity directly affects the material's temperature under adiabatic conditions, thus quantifying the magnetic entropy change associated with MCE which can be achieved by measuring magnetization and temperature changes under external magnetic field.[5–7]

Such MCE is a promising alternative for sustainable development. It enables temperature change not only without the use of fluorinated refrigerants such as chlorofluorocarbon (CFC), which have caused approx. 10% of the atmosphere's ozone depletion, but also higher energy efficiency than using the refrigerants.[8–10] However, the commercialization of MCE requires the materials to display high entropy change near room temperature due to many practical applications working near room temperature such as air conditioning and refrigerators.[11,12] In other words, the Curie temperature should exist near room temperature where magnetic phase transition appears between ferromagnetic and paramagnetic.

Currently, materials including rare-earth elements, such as Gd, are used due to their substantial MCE capabilities near room temperature, but their high cost (approx. 300

USD/kg) poses a big challenge for wide applications.[13–15] This motivates a search for new materials based on Earth-abundant elements to make them cost-effective and that can deliver high MCE performance near room temperature.[16] As part of this effort, alloying with relatively low-cost elements such as Si and Ge into Gd, and various perovskite oxides have been reported.[14,17–19] Gd alloys containing transition metals of Fe, Ni, Cr etc. have been extensively studied for their MCE properties.[20–22] However, research on outstanding MCEs of low-cost transition metals alone remains limited. Also, while nanoparticles exhibit generally enhanced MCE performance than bulk due to superparamagnetic, investigating the MCE of nanoparticles synthesized from such bulk by top-down approach is seldom.[23,24]

Herein, we synthesized and investigated the MCE performances of $Fe_{47.5}Ni_{37.5}Mn_{15}$ (at.%) in both bulk and nanoparticle forms. This specific composition, characterized by a Curie temperature near room temperature and significantly lower cost compared to rare metal-based candidates, presents a promising option for commercialization.[25] The $Fe_{47.5}Ni_{37.5}Mn_{15}$ bulk shows superior MCE performance than Gd for cost effectiveness, analyzed by microstructure and composition analysis. Additionally, the impact of particle size reduction on enhancing MCE performance was examined along with in-depth analysis of the impurities and composition.[26] How the composition was varied and the impurities from the laser ablation solution were embedded in $Fe_{47.5}Ni_{37.5}Mn_{15}$ nanoparticles produced by pulsed-laser ablation in ethanol is discussed.

**Experimental Section**

Bulk $Fe_{47.5}Ni_{37.5}Mn_{15}$ alloy was produced by melting 14 g of Fe (99.99% purity, iNexus),

11 g of Ni (99.95% purity, Avention), and 4 g of Mn (99.9% purity, Avention) metals using Ar-arc plasma. The as-cast alloy was then homogenized at 1050°C for 120 hours in Ar atmosphere and quenched in water. $Fe_{47.5}Ni_{37.5}Mn_{15}$ nanoparticles were synthesized by using pulsed-laser ablation in liquids method.[27] During laser ablation, a high-energy laser beam was focused on the cast bulk target immersed in ethanol, resulting in a colloidal solution from which the nanoparticles were collected by centrifugation and dried in a vacuum desiccator. The laser ablation experiments were carried out using a nanosecond pulsed laser (EdgeWave GmbH) with a wavelength of 1064 nm, pulse duration of 7 ns, repetition rate of 5 kHz and pulse energy of around 30 mJ. A continuous flow chamber was used with ethanol pumped at a flow rate of 50 ml/min.

Magnetization properties including the Arrott plot and isothermal magnetization curve, were characterized using a vibrating sample magnetometer (VSM, MPMS3-Evercool, Quantum Design).[28,29] Microstructural characterization was performed using scanning electron microscopy (SEM, SU8200 Hitachi and Merlin Zeiss) with 15kV acceleration voltage, utilizing backscattered electron (BSE), an energy-dispersive x-ray spectroscopy (EDX), and electron backscatter diffraction (EBSD). EBSD maps with dimensions of 2100 x 1400 µm$^2$ were acquired, with a step size of 1.5 µm EBSD maps were analyzed with the orientation imaging microscopy (OIM) software, v. 8.6. Only pixels with confidence index (CI) ≥ 0.1 were used for analysis. X-ray diffraction (XRD, smart lab, Rigaku and d8 advance, Bruker) analysis was conducted by Co kα (wavelength: 0.178897 nm) source and Cu kα (wavelength: 0.154059 nm) with 45 kV and 200 mA settings. The Rietveld refinement was performed by using the Fullprof package. The scanning transmission electron microscope (STEM, Titan Themis 300, Thermo Fisher) images and STEM-EDS maps were collected using high-angle annular dark field (HAADF) at an acceleration voltage of 300 kV.

Both bulk and nanoparticle samples were prepared into atom probe tomography (APT) specimens using a standard lift-out procedure with a focused ion beam (FIB) (FEI Helios 600 and Helios G5 CX).[30,31] Prior to nanoparticle APT specimen preparation, the nanoparticles were co-electrodeposited with a Co layer, which has a similar field evaporation.[32–34] The coating solution was prepared using 48g of $CoSO_4·7H_2O$ (98% purity, DAEJUNG), 12g of $CoCl_2·6H_2O$ (97% purity, DAEJUNG), and 9g of $H_3BO_3$ (99.5% purity, JUNSEI) in 200 mL of de-ionized water, following a previously established electrodeposition procedure.[35] APT was performed using both LEAP 4000XR and 5000XR (CAMECA) instruments under the following conditions: stage temperature at 50K, a detection rate of 0.5%, laser-power energy of 60 pJ, and pulse rates of 200 kHz.

**Result & discussion**

Figure 1a displays the magnetization curve of $Fe_{47.5}Ni_{37.5}Mn_{15}$, which exhibits characteristics of a second-order magnetization transition, indicated by the absence of a sharp slope under 1 T field (see the 1st derivative curve in the inset in Figure 1a). $T_c$ is approximately 300 K, as this is where the change in magnetization with respect to temperature is maximized. Figure 1b presents the isothermal magnetization curve, demonstrating a paramagnetic property below 375 K as indicated by a linear increase under an external magnetic field. The Arrott plot in Figure 1c reveals the order of the magnetic transition: a negative slope indicates a first-order transition, while a positive slope denotes a second-order transition. Additionally, for temperatures below $T_c$, in the ferromagnetic region, the curve bends upward at the end point, whereas for temperatures above $T_c$, the curve bends downward (e.g. paramagnetic region). The temperature at which the plot displays a linear relationship through the origin is regarded as $T_c$.[28] In Figure 1c, a linear behavior observed near 300 K suggests a magnetic phase transition from ferromagnetic to paramagnetic. The change in magnetic entropy increases with the intensity of the applied magnetic field as shown in Figure 1d. $T_c$ is also inferred from the temperature corresponding to the maximum entropy change, which is approximately 300 ± 5 K under 1 T and near 327 K under 5 T. At a higher magnetic field, the spin alignment by external magnetic field is well maintained at elevated temperatures, as thermal randomization of spins is suppressed under these conditions.

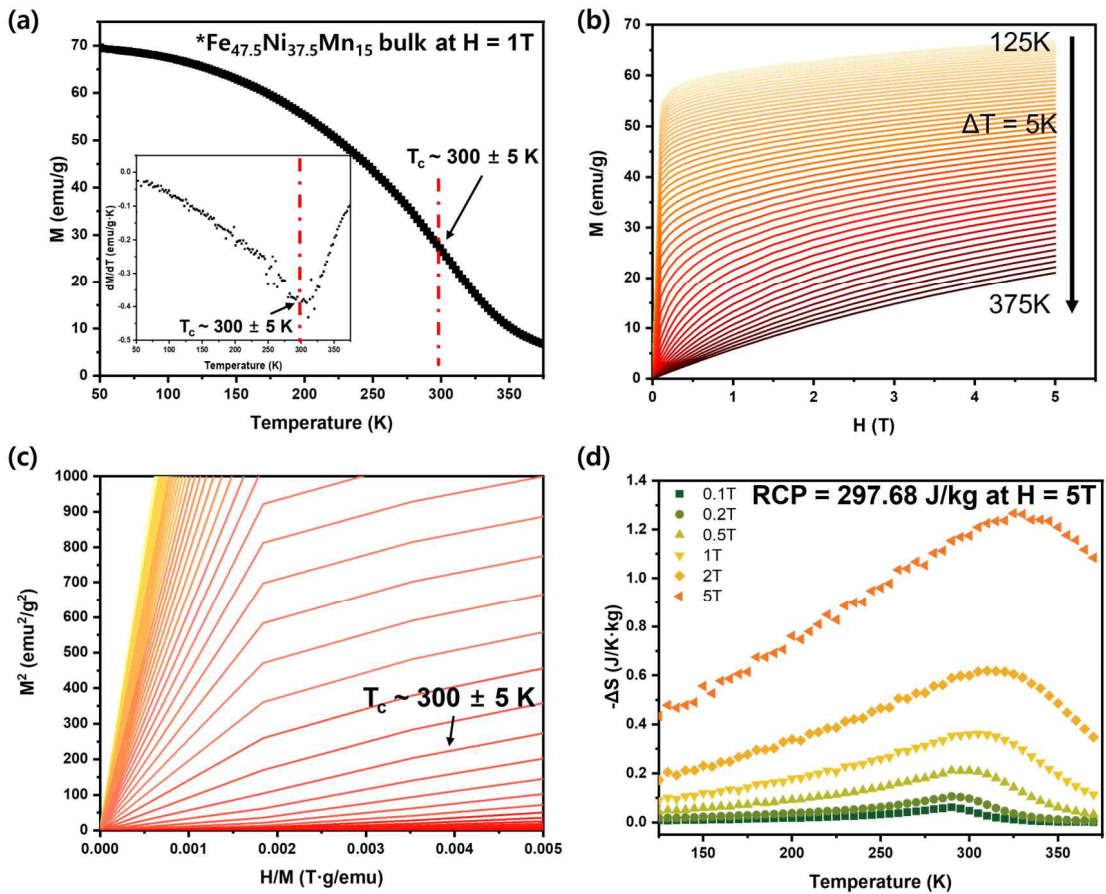

**Figure 1.** (a) Magnetization curve versus temperature from 50 K to 375 K under H = 1 T and inset image is the first order derivative of the magnetization curve. (b) Isothermal magnetization profile of bulk $Fe_{47.5}Ni_{37.5}Mn_{15}$ from 125 K to 375 K in 5 K intervals. (c) Arrott plot of the bulk $Fe_{47.5}Ni_{37.5}Mn_{15}$ near the origin from 125 K to 375 K in 5 K intervals. The full range plot is attached in Figure S1. (d) Temperature versus magnetic entropy change curve under different magnetic fields 0.1, 0.2, 0.5, 1, 2, and 5 T.

The relative cooling power (RCP), an indicator of MCE capability, is calculated by multiplying the maximum entropy change ($-\Delta S_{max}$) by the temperature difference at half maximum ($\delta T_{FWHM}$), providing comprehensive measure of the cooling efficiency of a material.[36] Under an applied external magnetic field of 5 T, the RCP of the $Fe_{47.5}Ni_{37.5}Mn_{15}$ bulk is expressed as 297.68 J/kg, compared to 410 J/kg for Gd.[37] Despite the lower performance of $Fe_{47.5}Ni_{37.5}Mn_{15}$, considering the price of its components (Fe: 0.31 USD/kg, Ni: 22 USD/kg, Mn: 0.25 USD/kg), $Fe_{47.5}Ni_{37.5}Mn_{15}$ is significantly more economical than Gd (approx. 300 USD/kg). With a cost approximately 35.7 times lower, the alloy offers comparable performance being around 25.8 times more efficient in terms of RCP per unit cost.[38,39]

To identify factors contributing to MCE performance, the microstructure of the bulk material was examined. Figures 2a-2c show that the average grain size equivalent diameter is 75 ± 57 µm. Hu et al reported that smaller grain size of $La_{0.7}Ce_{0.3}Fe_{11.6}Si_{1.4}C_{0.2}$ from 20 µm to 50 µm increases the magnetization change due to hysteresis reduction by easy heat conduction, whereas coarser grain size from 90 µm to 120 µm increases the change of magnetic entropy. The grain size of $Fe_{47.5}Ni_{37.5}Mn_{15}$ bulk is also widely distributed as in the above case. Thus, the high RCP of $Fe_{47.5}Ni_{37.5}Mn_{15}$ bulk would be contributed from the interplay between hysteresis reduction and the increase of magnetic entropy change.[40] Figures 2b and

S2a exhibit randomly oriented grains as the maximum value (here, 1.82 multiple intensity of random distribution, mrd) denoted the extent of specific grain orientation in IPF is less than 3 mrd.[41] Textured materials generally show improved MCE performance than randomly orientated grains due to increased magnetization on specific orientation.[42,43] For instance, Fe-3.85Si wt% alloy represents stronger magnetization in [100] direction than in [110] and [111] directions, conveying that a specifically oriented crystal structure can increase the MCE. Thus, the MCE performance of $Fe_{47.5}Ni_{37.5}Mn_{15}$ bulk does not seem to be attributed to texture.

The XRD pattern in Figure 2d displays the peaks at 2θ = 51.2°, 59.7°, 89.5°, 110° and 119°, indicating face-centered cubic (FCC) crystal structure without any noticeable peaks from impurity phases within the detection limit of the technique. The FCC Ni-Fe binary alloy typically exhibits higher magnetization compared to pure Fe due to the contribution of Fe moment, probably resulting in superior MCE performance of bulk material.[44] Figures 2d and S3 confirm a random distribution of Fe, Ni, and Mn atoms without segregation or clustering as the alloy was prepared using a homogenization process. In Figure S3, TiN, VN, MnS inclusions that commonly exist in Fe-based alloy were observed in the matrix despite not being detected in XRD.[45–47] Yet, they do not seem to influence the magnetization owing to the absence of any plateau in the change of magnetic entropy versus temperature (Figure 1d).[48] The 3D elemental atom maps in Figure 2e show a homogenously distributed composition, as indicated by calculated Pearson coefficients of 0.07 for Fe, 0.08 for Ni, and 0.04 for Mn atoms. Despite the low Pearson coefficient, there may be subtle short-range order (SRO) in the bulk that cannot be seen due to the resolution limit, which could cause non-linear isothermal magnetization curve above Curie temperature as Souza et al. reported.[49–

51] The measured composition of Fe (48.4 at.%), Ni (36.1 at.%), and Mn (15.2 at.%) closely aligns with the target composition $Fe_{47.5}Ni_{37.5}Mn_{15}$. Since the different spin exchange interactions between atoms by composition variation could provoke unwanted changes in $T_c$, this well-matched composition as reported gives rise to $T_c$ located near room temperature.[22,25]

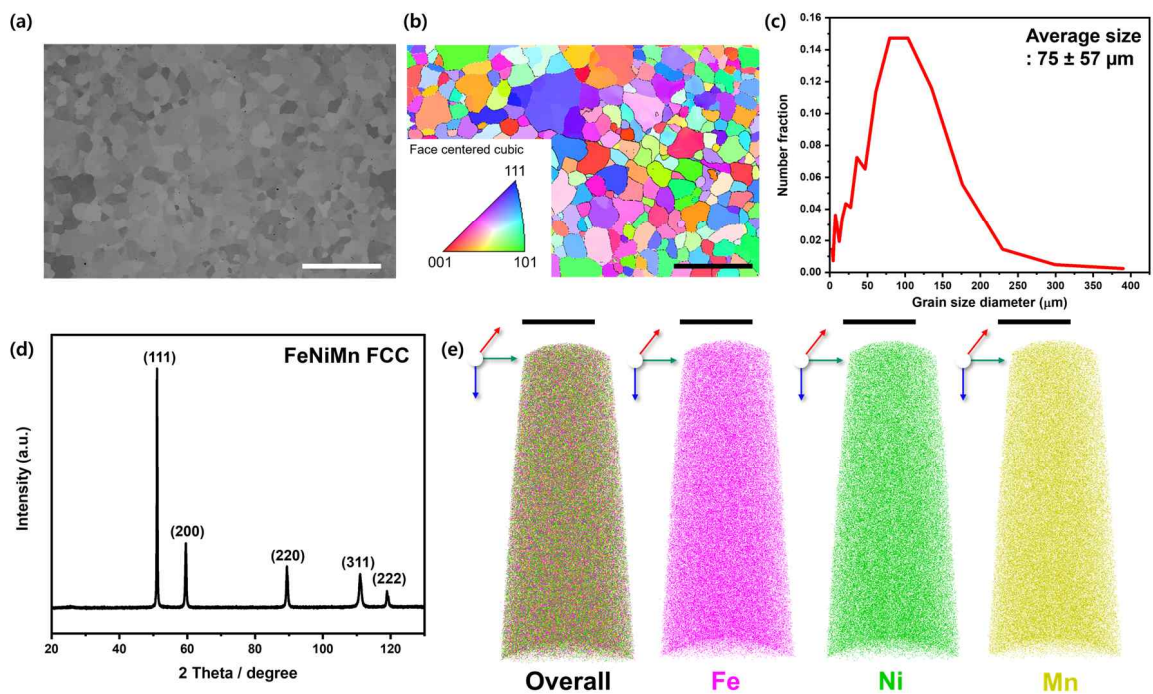

**Figure 2.** (a) BSE-SEM image of bulk $Fe_{47.5}Ni_{37.5}Mn_{15}$. (b) The inverse pole figure map from EBSD indicating the crystallographic direction of each grain along the normal direction (perpendicular to the analyzed surface). Scale bars in (a) and (b) are 500 μm. (c) Grain size distribution of bulk $Fe_{47.5}Ni_{37.5}Mn_{15}$ (d) XRD of bulk $Fe_{47.5}Ni_{37.5}Mn_{15}$, (e) APT reconstruction and atom maps of Fe, Ni and Mn. Scale bars in (e) are 50 nm.

To further enhance the cooling efficiency of these materials, reducing them to nanoscale sizes offers significant advantages including faster response times and more efficient heat exchange during the magnetocaloric process.[23,52] Additionally, dispersing nanoparticles in a coolant suspension allows for compact and direct heat transfer during circulation.[22,53] In particular, when a material's size is reduced to the nanoscale, particle-size effects become significant in magnetic materials.[26] Specifically, superparamagnetic effects become prominent, leading to the paramagnetic behavior of originally ferromagnetic or ferrimagnetic materials due to their reduced size. This enhances the efficiency of MCE due to higher changes of magnetic entropy and fast heat conduction.[54,55] However, if the particles become too small, their magnetic moments and blocking temperature, which is a magnetic phase transition from superparamagnetic to ferromagnetic or ferrimagnetic, decrease, leading to a reduced change in entropy near room temperature, which ultimately diminishes the efficiency of the MCE.[56–58]

Here, nanoparticles were synthesized via pulsed-laser ablation in liquids method. This approach allows for facile one-step production of nanoparticles from bulk metal alloys. Besides, this synthesis method is suitable for mass production with no necessity for the addition of ligands nor reducing agent for stabilization, and can hence produce high-purity nanoparticles.[27,59–62] However, in Figure 3a, the nanoparticles present a lower magnetization value (~ 34 emu/g at 50 K) than the bulk material (~70 emu/g at 50 K), which could be correlated to the existence of secondary phases. Additionally, a distinct straight line is not observed up to 375 K, indicating no magnetic phase transition (e.g. $T_c$ is above 375 K). Likewise, Figures 3b and 3c suggest that the maximum entropy change occurs at a temperature above 375 K as the Arrott plot shows no straight line up to this temperature. Presumably, the Curie temperature is calculated as 477 K, as explained in supplementary information (Figure S4). As shown in Figure 3d, the peaks around 325 K were detected, possibly coming from the non-uniform composition distribution between nanoparticles as discussed below. The maximum entropy change peaks are not observed up to 375 K, indicating relatively poor RCP performance near room temperature.

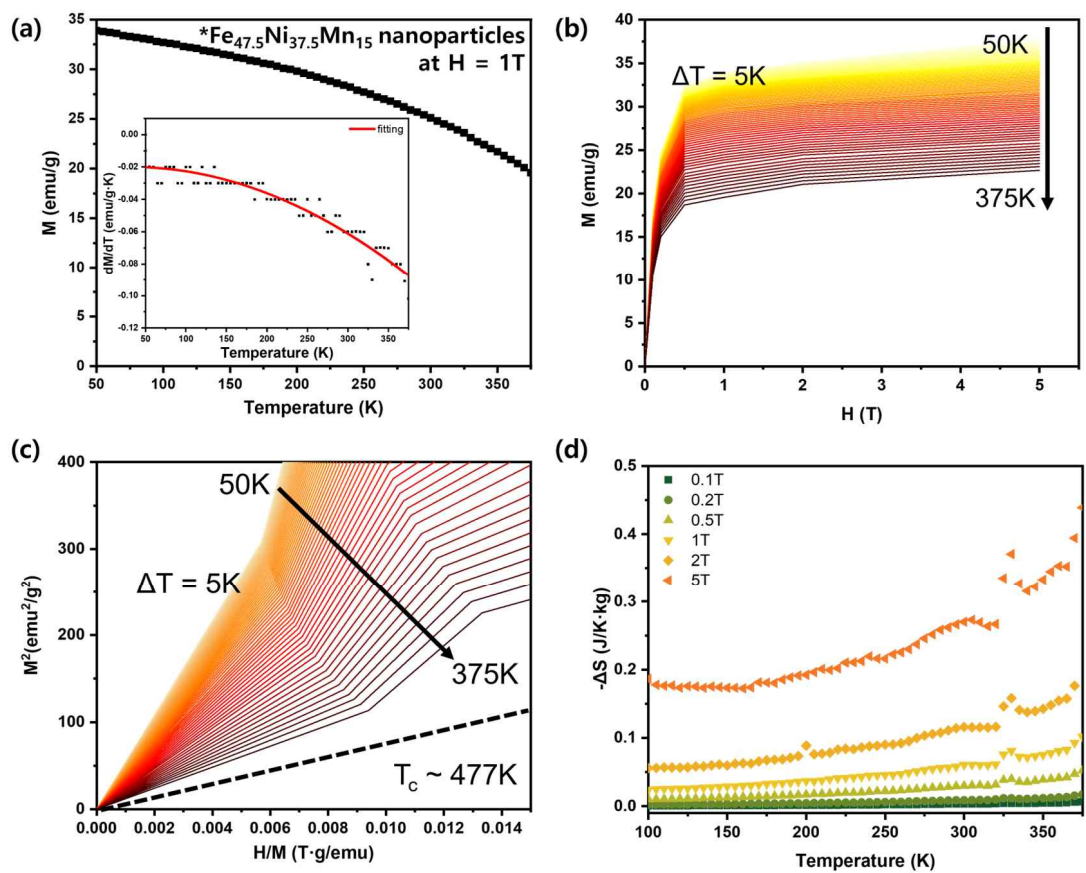

**Figure 3.** (a) Magnetization curve of $Fe_{47.5}Ni_{37.5}Mn_{15}$ nanoparticles versus temperature under H = 1 T and inset image is the first order derivative of the magnetization curve. (b) Isothermal magnetization profile of $Fe_{47.5}Ni_{37.5}Mn_{15}$ nanoparticles from 50 K to 375 K in 5 K intervals. (c) Arrott plot of $Fe_{47.5}Ni_{37.5}Mn_{15}$ nanoparticles near the origin from 50 K to 375 K in 5 K intervals. The full range is attached in Figure S5. (d) Temperature versus magnetic entropy change curves under different magnetic fields 0.1, 0.2, 0.5, 1, 2, and 5 T.

To investigate the reason behind the unexpected increased $T_c$ for the nanomaterials, microstructural and compositional analyses were conducted. In Figure 4a, the Fe, Ni, and Mn exist within nanoparticles, but carbon was observed both inside and surrounding the nanoparticles, interconnecting between them. As shown in Figures S6a and S6b, the laser-ablated nanoparticles were identified to be spherical but agglomerated. interconnecting them While this agglomeration broadens the magnetic entropy change near the blocking temperature, it does not contribute to an increase in $T_c$.[63,64] The average nanoparticle size observed by TEM is measured as 9.5 ± 7.0 nm in Figure 4b, presenting the small enough particle size to show superparamagnetic as most superparamagnetic occur at 3 nm to 50 nm nanoparticle size.[65] In Figure 4c, the Rietveld refinement of XRD profile displays peaks around 2θ = 42.7°, 49.7°, 72.2°, and 88.4° corresponding to the FCC crystal structure (space group: Fm-3m, No.16). Concerning the Scherrer Equation D = Kλ/βcosθ, where D is average crystallite size, K is the Scherrer constant (usually 0.9), λ is X-ray wavelength, β is FWHM (full width at half maximum) of the peak, θ is the peak position, the average size is calculated as 18.3 nm at 2θ = 42.7°. Albeit with a higher value than derived from TEM likely due to the Scherrer constant affected by the shape and distribution of nanoparticles, this value also shows small enough to show superparamagnetic.[66,67] Additional peaks at 2θ = 39.3°, 41.5°, 44.6°, 58.5°, and 77.8° could be attributed to $Ni_3C$ (space group: R-3c, No.167) phase formed due to carbon ingress during laser ablation in ethanol medium resulting from solvent disintegration.[68–71] According to Rietveld refinement, the FCC phase and $Ni_3C$ impurity phase are about 15 vol.% and 85 vol.%, respectively. Although crystalline $Ni_3C$ is known as nonferromagnetic behavior due to strong hybridization between Ni and C, the magnetization of nanoparticles at 50K is higher than 15% of the magnetization of the bulk at 50K. This could originate from the local Ni-rich region in the $Ni_3C$ structure acting as ferromagnetic Ni.[72,73]

Figure 4d shows a 3D reconstruction map obtained by APT, illustrating the nanoparticles embedded into the Co matrix. Although Co is not expected to be inside the nanoparticle, Co was observed since distinguishing its signal from NiH signal, possibly formed by residual $H_2$ in The APT analysis chamber or through similar deposition potential with hydrogen, is challenging.[74,75] The $Ni_3C$ phase expressed by 12 at. % Ni iso-surface was composed of Ni 76.1 ± 2.8 at.% and C 23.9 ± 2.0 at.%, supporting XRD result. Within 5 at.% Fe iso-surface, the nanoparticle composition was measured as Fe 35.5 at.%, Ni 29.5 at.%, Mn 11.8 at.%, C 23.2 at.% in region of interest (ROI) #1 and Fe 46.6 at.%, Ni 28.0 at.%, Mn 2.81 at.%, C 22.6 at% in ROI #2, indicating not only the transition metal ratio difference between nanoparticles but also heavily ingressed carbon in both nanoparticles. This carbon amount inside both nanoparticles almost reaches the carbon solubility limit of 25 at.% in FCC structure.[76] The composition difference can arise from the discrepancy of latent heat evaporation of each element during laser ablation. Jakobi et al reported that laser-ablated nanoparticles from Sm-Co alloy display the composition difference from the target due to heat evaporation between elements.[77] Although they mentioned nanoparticles from Fe-Ni alloy exhibit relatively similar composition with target owing to similar latent heat (Ni:370.4 kJ/mol Fe:349.6 kJ/mol), the lower heat evaporation of Mn (266 kJ/mol) in our Fe-Ni-Mn ternary system might generate composition deviation of nanoparticles, possibly leading to the degradation of MCE performance near room temperature by increase of $T_c$ due to changing spin interaction between transition metals.

Figures 4e and 4f show that the carbon impurity exists inside the nanoparticles and around the shell as observed by TEM.[71] According to the literature, ethanol can be decomposed at 1000 K and laser ablation has enough energy to elevate the temperature up to 4000 K.[78,79] The products of decomposed ethanol, such as $CH_4$ and $C_2H_4$ gases used for the carburization of steel, were diffused into the nanoparticles due to high thermal energy.[80] Furthermore, organic radicals from the thermal decomposition can react with metal ions, possibly forming a bond at the surface, which can be correlated to the carbon near the nanoparticle surface as identified by TEM and 1-D concentration profiles.[81] Consequently, the nanoparticle produced by pulsed-laser ablation in ethanol shows significant carbon impurity both inside and outside, influencing the increase of $T_c$ as even a small amount of carbon considerably impacts the variation of $T_c$.[22,82,83] Thus, selecting a suitable solvent medium to reduce carbon content during synthesis is crucial to achieving a room temperature $T_c$.[84]

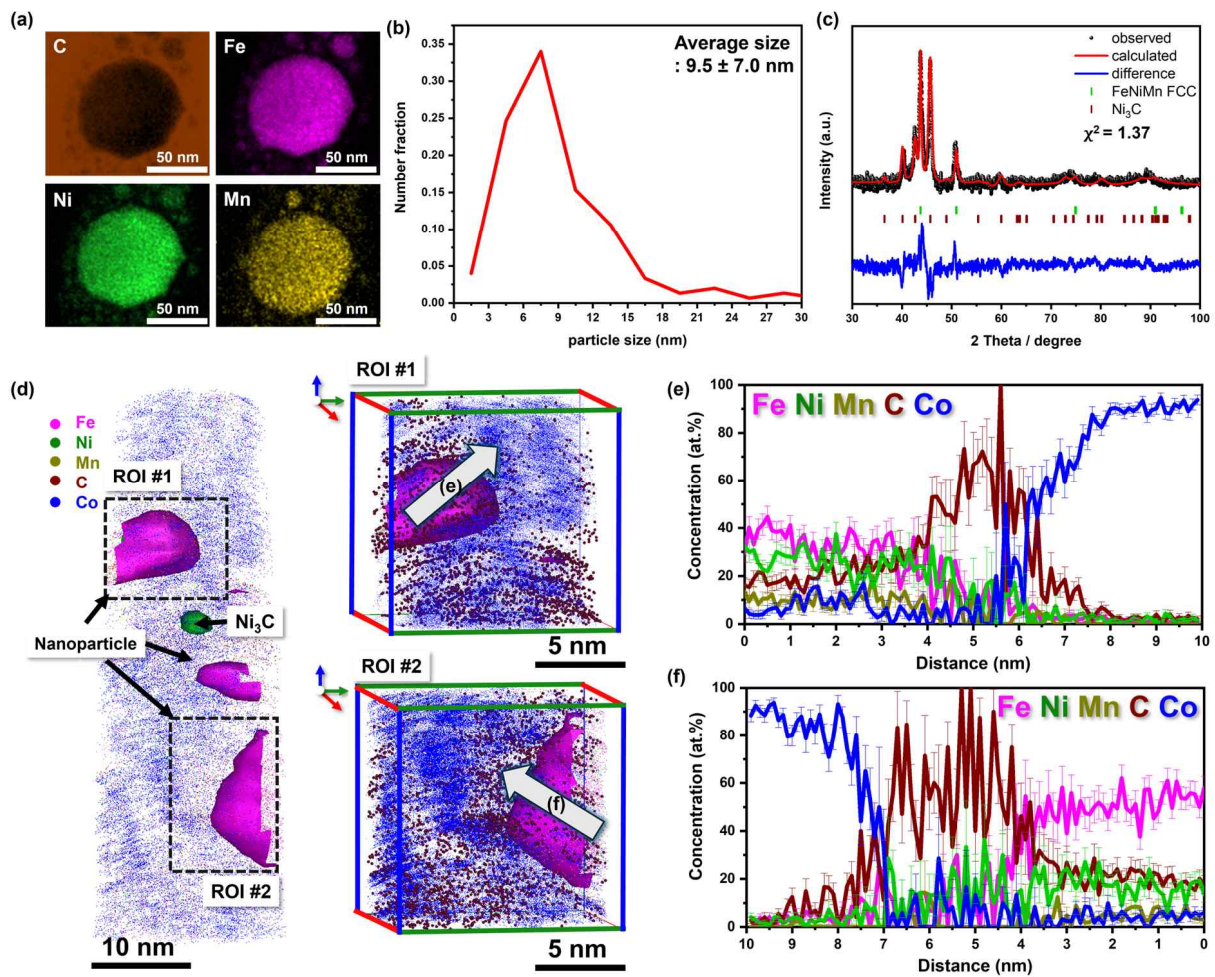

**Figure 4.** (a) STEM-EDS maps of $Fe_{47.5}Ni_{37.5}Mn_{15}$ nanoparticles. (b) Particle size distribution of nanoparticles. (c) Rietveld refinement of X-ray diffractogram of nanoparticles. (d) APT reconstruction of the nanoparticles embedded in the Co matrix and two regions of interest near nanoparticles. The 1D concentration profile with a bin width of 0.1 nm in the region of interest ($\Phi 5 \times 5 \times 10$ nm³) of (e) ROI #1 and (f) #2.

It is crucial to design (nano)-materials that exhibit MCE at room temperature and are cost-effective for commercialization. In that sense, transition metals like Fe, Mn, and Ni are promising for potential commercial MCE materials. This study compared the MCE performance of bulk $Fe_{47.5}Ni_{37.5}Mn_{15}$ and the nanoparticles near room temperature. Unexpectedly, the nanoparticles known as better MCE performance owing to super-paramagnetic, exhibit poorer MCE performance near room temperature since carbon ingress from laser ablation increases $T_c$. Thus, the optimized top-to-bottom or bottom-to-top synthesis steps to prevent impurity ingress from causing unwanted effects is essential for achieving the desired MCE performance of nanoparticles.

**Conclusion**

A cost-effective $Fe_{47.5}Ni_{37.5}Mn_{15}$ alloy composed of transition metals was designed and its magnetic properties were characterized, demonstrating a promising MCE with an outstanding RCP of 297.68 J/kg at room temperature. $Fe_{47.5}Ni_{37.5}Mn_{15}$ nanoparticles were synthesized via pulsed-laser synthesis approach to enhance MCE, aiming to leverage superparamagnetic effects. However, the nanoparticles exhibited a MCE shifted to high temperatures (~477 K), due to carbon ingress during laser ablation, as detected via APT. Although the $Fe_{47.5}Ni_{37.5}Mn_{15}$ alloy shows potential for cost-effective

MCE applications, it seems to be necessary to optimize the nanoparticle synthesis method, pulsed-laser ablation, for better MCE at room temperature.

**Supporting information**

Supporting information (SI) will be uploaded online.

**Acknowledgement**

This work was supported by the Technology Innovation Program (or Industrial Strategic Technology Development Program-Establishment of TiAl Material Characteristics DB for Aero Engines) (RS-2024-00431836) funded By the Ministry of Trade, Industry & Energy (MOTIE, Korea) and the Korea Institute of Energy Technology Evaluation and Planning (KETEP) (No. RS-2024-00401917). This work was also supported by the Hyundai Motor Chung Mong-Koo Foundation. A.A. El-Zoka acknowledges the support of EPSRC grant reference number : EP/V007661/1

**Conflict of interest**

The authors declare no conflict of interest.

# Magnetocaloric effect of $Fe_{47.5}Ni_{37.5}Mn_{15}$ bulk and nanoparticles: A cost-efficient alloy for room temperature magnetic refrigeration

Chang-Gi Lee[a], Varatharaja Nallathambi[b,c], TaeHyeok Kang[d], Leonardo Shoji Aota[c], Sven Reichenberger[b], Ayman El-Zoka[e], Pyuck-Pa Choi[d], Baptiste Gault[c,e], Se-Ho Kim[a,c,*]

[a] Department of Materials Science & Engineering, Korea University, Seoul, 02841, Republic of Korea

[b] Technical Chemistry I and Center for Nanointegration Duisburg-Essen (CENIDE), University of Duisburg-Essen, 45141 Essen, Germany

[c] Max-Planck-Institut für Eisenforschung GmbH, Max-Planck-Straße 1, 40237 Düsseldorf, Germany

[d] Department of Materials Science & Engineering, Korea Advanced Institute of Science and Technology, Daejeon, 34141, Republic of Korea

[e] Faculty of Engineering, Department of Materials, Imperial College London, London, SW7 2AZ, United Kingdom

*corresponding author: sehonetkr@korea.ac.kr

**Thermodynamic principle of magnetocaloric effect**

The total entropy of a magnetic material ($S_{tot}$) is generally expressed as follows:

$$S_{tot} = S_e + S_l + S_m \quad \text{Eq. 1}$$

where $S_e$, $S_l$, and $S_m$ denote the entropy contributed from electrons, lattice, and magnetism, respectively.[1,2] For simplicity, the $S_e$ and $S_l$ are usually assumed to vary only with temperature.[1,3] When an external magnetic field is applied to the material, the magnetic entropy is decreased because of the alignment of the spin orientation. Since the total entropy remains constant under reversible conditions, the other entropy term should increase with a rise of temperature. The change in the magnetic entropy ($\Delta S_m$) based on the above assumptions, emerges as a critical metric for evaluating the MCE, and is defined by the following equation,

$$\Delta S_m(T, B) = \frac{C_B(T,B)}{T} dT + \left(\frac{\partial S_m(T,B)}{\partial B}\right)_T dB \quad \text{Eq. 2}$$

where $C_B(T, B)$, T, and B are the heat capacity, temperature, and external magnetic field, respectively.[4,5] The effectiveness of MCE is often characterized by the magnitude of $\Delta S_m$ which indicates how much entropy changes during the application or removal of the external magnetic field.[2]

From Maxwell's relation, $\Delta S_m(T, B)$ under isothermal conditions can be expressed as Eq. 3, as symmetry second derivative of state function in thermodynamics is equal

$$\Delta S_m(T, B) = \int_{B_1}^{B_2} \left(\frac{\partial M(T,B)}{\partial T}\right)_B dB \quad \text{Eq. 3}$$

where $B_1$, $B_2$, and $M(T, B)$ represent the initial/final magnetic fields and the magnetization, respectively.[6] From this equation 3, the change in magnetic entropy of the material is measured by integrating the derivative of magnetization by temperature when magnetic field is varied from $B_1$ to $B_2$.

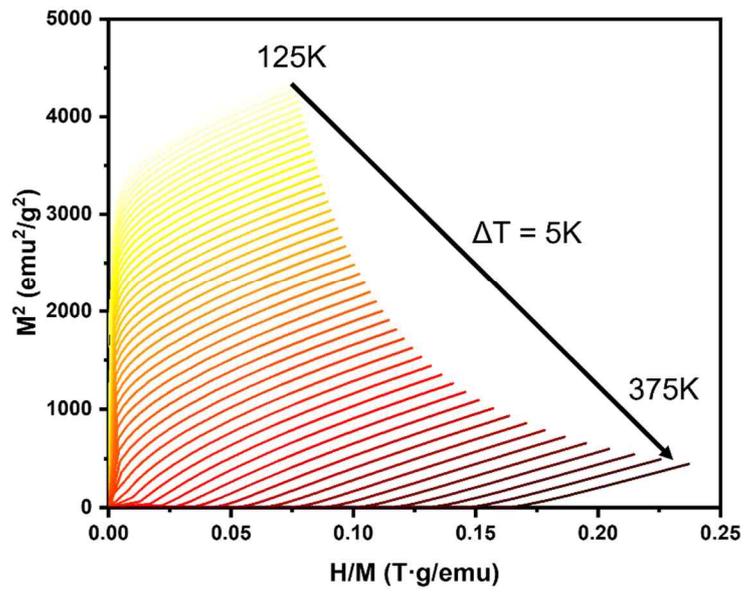

**Figure S1.** Arrott plot of $Fe_{47.5}Ni_{37.5}Mn_{15}$ bulk from 125 K to 375 K in 5 K intervals.

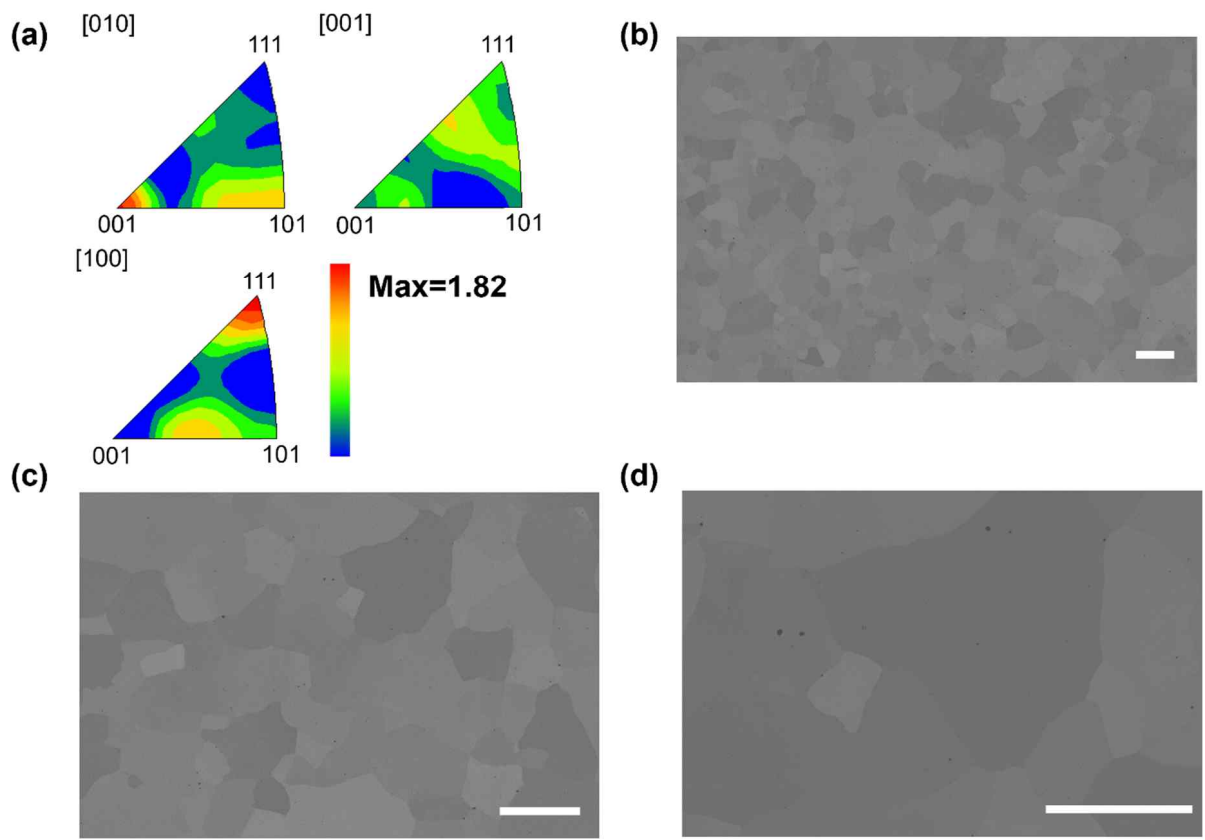

**Figure S2.** (a) Inverse pole figure (IPF) maps. BSE-SEM images of $Fe_{47.5}Ni_{37.5}Mn_{15}$ bulk with magnifications (b) x50 (c) x100 (d) x250. Scale bars stand for 100 μm.

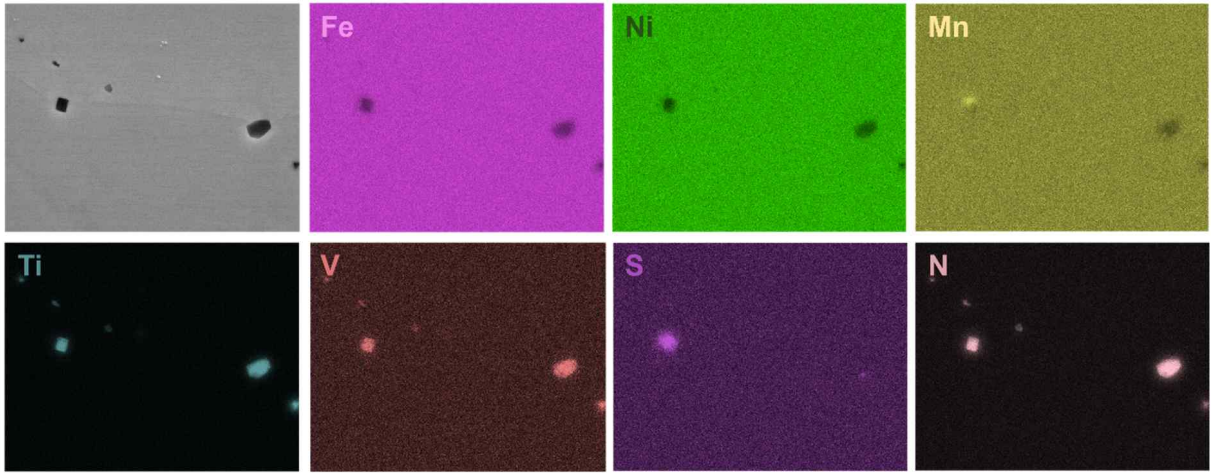

**Figure S3.** SEM-EDS mappings of $Fe_{47.5}Ni_{37.5}Mn_{15}$ alloy

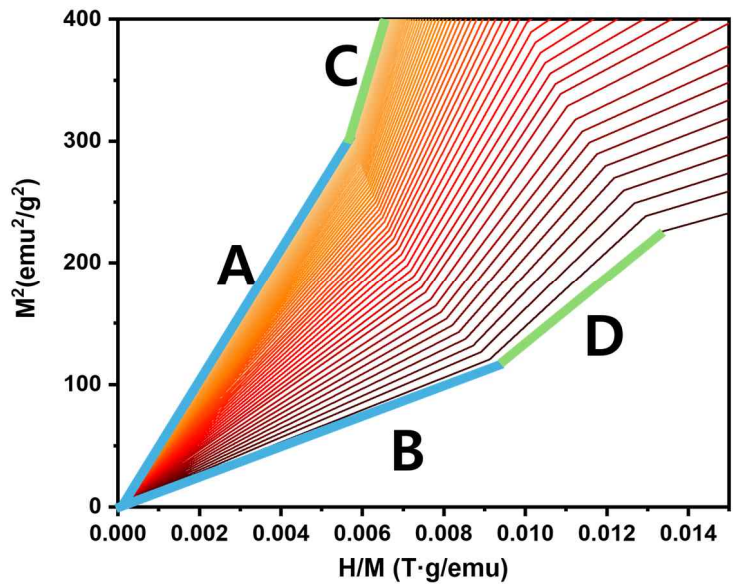

**Figure S4.** Enlarged Arrott plot of nanoparticles near the origin.

The Curie temperature of nanoparticles located above room temperature can be deduced by calculating as follows. The Curie temperature is graphically obtained when the slope of the straight line between the nearest point from the origin and the origin becomes equal to the slope of the straight line between the first and second nearest point from the origin.

When the temperature is increased from the initial ($T_1$, here 50 K) to the final temperature ($T_2$, here 375 K), it was assumed that the decrease rate of the slope of the straight line is linear. The slope of each straight line in Figure S6 is A, B, C, and D, respectively. Here, A is the slope of the line between the origin and the first nearest point from the origin at the initial temperature, and B is the slope of the line between the origin and the first nearest point from the origin at the final temperature. C is the slope of the straight line between the first and second nearest point at initial temperatures, and D is the slope of the straight line between the first and second nearest point at final temperatures. The Curie temperature is calculated using the below equation:

$$x\,(K) = \frac{(C-A)(T_2-T_1)}{(-A+B+C-D)} + T_1$$

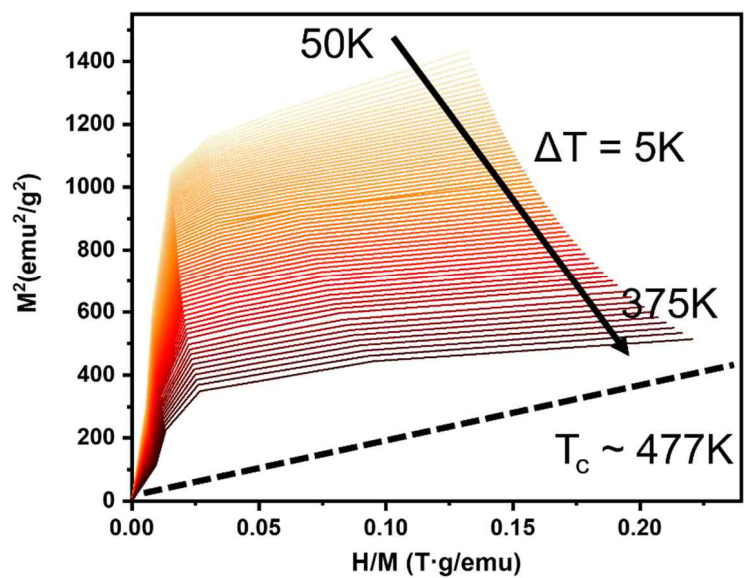

**Figure S5.** Arrott plot of $Fe_{47.5}Ni_{37.5}Mn_{15}$ nanoparticles from 50 K to 375 K in 5 K intervals.

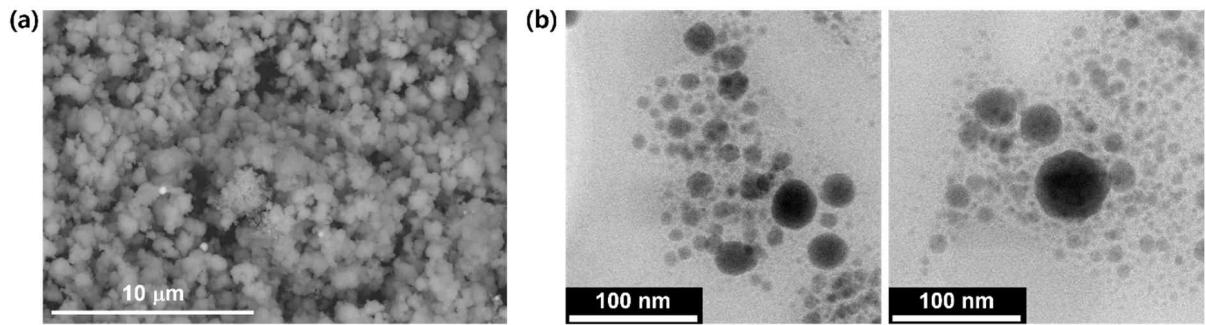

**Figure S6.** (a) SEM image and (b) HAADF-STEM images of $Fe_{47.5}Ni_{37.5}Mn_{15}$ nanoparticles